\documentclass[a4paper,10pt]{article}
\usepackage{amssymb}
\usepackage{latexsym,graphicx,epsfig,psfrag,here}
\usepackage{subfigure}
\usepackage{textcomp}

\begin{document}
\begin{titlepage}
\begin{flushright}
{UAB-FT-725}
\end{flushright}
\vskip 1.5cm

\begin{center}
{\LARGE \bf Dispersive representation of the pion vector form factor in
$\tau\to\pi\pi\nu_\tau$ decays}
\\[50pt]
{\sc  D.~G\'omez Dumm$^{1}$,
P.~Roig$^{2}$}

\vspace{1.4cm} ${}^1$ IFLP, CONICET $-$ Dpto. de F\'{\i}sica,
Universidad Nacional de La Plata,  \\ C.C.\ 67, 1900 La Plata, Argentina. \\[15pt]
${}^2$ Grup de F\'{\i}sica Te\`orica, Institut de F\'{\i}sica d'Altes Energies,
Universitat Aut\`onoma de Barcelona, E-08193 Bellaterra, Barcelona, Spain.\\[10pt]
\end{center}

\vfill

\begin{abstract}
We propose a dispersive representation of the charged pion vector form
factor that is consistent with chiral symmetry and fulfills the constraints
imposed by analyticity and unitarity. Unknown parameters are fitted to the
very precise data on $\tau^-\to\pi^-\pi^0\nu_\tau$ decays obtained by Belle,
leading to a good description of the corresponding spectral function up to a
$\pi\pi$ squared invariant mass $s\simeq 1.5$~GeV$^2$. We determine the
$\rho(770)$ mass and width pole parameters and obtain the values of low
energy observables. The significance of isospin breaking corrections is also
discussed. For larger values of $s$, this representation is complemented
with a phenomenological description to allow its implementation in the new
TAUOLA hadronic currents.
\end{abstract}
\vspace*{3.0cm}

PACS~: 13.35.Dx, 12.38.-t, 12.39.Fe, 11.15.Pg, 13.40.Gp, 11.55.Bq
\\
\hspace*{0.45cm} Keywords~: Hadronic tau decays, Chiral Lagrangians,
$\rho(770)$ resonance, Electromagnetic form factors, Analytic properties of
S matrix. \vfill

\end{titlepage}

\section{Introduction}
Last years have witnessed a notorious progress in the knowledge of the
two-pion system, both from theoretical and experimental sides. In
particular, the high precision measurements of the charged and neutral pion
vector form factors performed at the flavour factories
BaBar~\cite{Aubert:2009fg}, Belle~\cite{Belle},
CMD-2~\cite{Akhmetshin:2006bx}, KLOE~\cite{Aloisio:2004bu} and
SND~\cite{Achasov:2005rg} have significantly improved the accuracy of
previous data.
As it is well known, theoretical predictions for these form factors cannot
be obtained analytically from first principles through standard
calculations, since this involves in general the hadronization of QCD
currents in a nonperturbative energy regime ($E\lesssim 1$ GeV). In order to
overcome this problem, one can rely on effective
models~\cite{Gounaris:1968mw, Pich:1989pq, Kuhn:1990ad} that intend to
describe the involved dynamics within a simplified scheme. In general, in
these models the form factors are required to satisfy the proper behaviour at
very low energies, and the effect of intermediate resonances is taken into
account through the inclusion of Breit-Wigner functions. In this way one can
obtain phenomenologically adequate hadronic matrix
elements~\cite{Decker:1992kj, Decker:1994af, Finkemeier:1995sr,
Kuhn:2006nw}, which have been included e.g.~in the standard Monte Carlo
Generator for tau decays, TAUOLA~\cite{TAUOLA}. However, in general these
models include several ad-hoc assumptions, and can even have problems of
consistency with QCD~\cite{Portoles:2000sr,Roig:2008xt}. In addition, even
when the agreement with experimental data can be very good, usually the
model parameters can be hardly related to the underlying strong interaction
theory.

Another possible approach is to consider just the symmetry properties of QCD
in order to build a general effective action adequate for the desired energy
regime. At very low energies $E \ll M_\rho$, where $M_\rho$ is the
$\rho(770)$ resonance mass, the approximate chiral symmetry of QCD allows to
build the effective quantum field theory known as Chiral Perturbation Theory
($\chi$PT)~\cite{ChPT}. The latter provides a successful description of the
low-energy phenomenology of strong and electroweak interactions, in which
hadronic observables are calculated through an expansion in powers of ratios
of momenta and masses of the lightest degrees of freedom (light pseudoscalar
mesons) over a chiral symmetry breaking scale, $4\pi F_\pi\sim1.2$ GeV.
However, for $E\sim M_\rho$ the expansion parameters become large, and new
degrees of freedom, namely the lowest-lying light-flavoured resonances,
become active. Even though in this regime there is no straightforward
expansion parameter, one can build an effective theory by considering an
expansion in powers of the inverse of the number of colours, $1/N_C$, with
the introduction of resonances as active fields in the effective action.
Indeed, it is found that this approach allows to describe satisfactorily
most salient features of meson phenomenology~\cite{Manohar:1998xv,
Pich:2002xy}, which suggests that the large-$N_C$ limit of QCD is a good
starting point to derive a chiral Lagrangian that includes
resonance fields \cite{Ecker:1988te, Ecker:1989yg, Cirigliano:2006hb}.

In this work we study one of the simplest hadronic observables, namely the
pion vector form factor $F_V^\pi(s)$, defined through
\begin{equation}\label{FF definition}
 \left\langle \pi^0\pi^-|\bar{d}\gamma^\mu
u|\emptyset\right\rangle\,=\,\sqrt{2}F_V^\pi(s)\left(p_{\pi^-}-p_{\pi^0}\right)^\mu\,,
\end{equation}
where $s\equiv q^2 \equiv (p_{\pi^-}+p_{\pi^0})^2$. For $s > 0$, this form
factor is probed by the decay $\tau^-\to\pi^-\pi^0\nu_\tau$, while in the
isospin symmetry limit it can be experimentally measured from
$e^+e^-\to\pi^+\pi^-$ ($s>0$) and elastic $e^-\pi^+$ scattering ($s<0$). The
analysis of $F_V^\pi(s)$ allows to increase our knowledge of the
hadronization of QCD currents in the intermediate energy region, where the
presence of meson resonances plays a crucial role. On the other hand, the
analysis of isospin breaking corrections to the form factors in tau decays
and $e^+e^-$ scattering~\cite{Davier:2002dy, Davier:2009ag,
Jegerlehner:2011ti, Benayoun:2011mm, Czyz:2013zga} is essential for the evaluation of the
hadronic contribution to the anomalous magnetic moment of the muon
$a_{\mu}$, which provides a stringent test of new
physics~\cite{Passera:2008jk}.

The theoretical analysis of the pion vector form factor has been addressed
by several authors in the last years. At very low energies, $F_V^\pi(s)$ has
been computed in $\chi$PT up to $\mathcal{O}(p^6)$~\cite{Gasser:1990bv,
Bijnens:1998fm, Bijnens:2002hp}. Then, to enlarge the domain of
applicability up to $\sim 1$ GeV, unitarization techniques~\cite{De
Troconiz:2001wt, Oller:2000ug} and dispersion relations have been
employed~\cite{Guerrero:1997ku, Pich:2001pj, Hanhart:2012wi}. Moreover, in
order to go beyond this energy region, the inclusion of the $\rho(1450)$
resonance~\cite{SanzCillero:2002bs} and even a tower of resonances, inspired
in the $N_C\to\infty$ limit~\cite{Dominguez:2001zu, Bruch:2004py}, have been
proposed. Our work is a sort of extension of those in
Refs.~\cite{Guerrero:1997ku,Pich:2001pj,Jamin:2006tk,Boito:2008fq}, in which
the authors analyze $\pi\pi$ and $K\pi$ vector form factors considering
$\mathcal{O}(p^4)$ expressions obtained from a chiral effective theory that
includes the dominant resonance exchange, followed by an Omn\`es-like
resummation of final state interactions. Our procedure is similar to that
proposed in Ref.~\cite{Boito:2008fq} for the $K\pi$ vector form factor: we
consider an $n$-subtracted dispersion relation for $F_V^\pi(s)$ in which the
input elastic phase shift $\delta_1^1(s)$ is taken from the effective
theory, resumming the chiral loops into the denominator of the
$\mathcal{O}(p^4)$ form factor. This ensures to fulfill unitarity and
analyticity constraints. It is seen that a phenomenologically good result is
obtained with three subtractions, hence our expression for $F_V^\pi(s)$
depends on four parameters: $M_\rho$, $F_\pi$, and two subtraction constants
$\alpha_1$ and $\alpha_2$ (the remaining subtraction constant is fixed by
the normalization of the form factor). These constants can be related to
chiral low-energy observables~\cite{Bijnens:1998fm, Bijnens:2002hp}, namely
the squared charged pion radius $\left\langle r^2\right\rangle^\pi_V$ and
the coefficients of $\mathcal{O}(s^2)$ and $\mathcal{O}(s^3)$ terms in the
chiral expansion, $c_V^\pi$ and $d_V^\pi$, respectively.

The above described approach is able to provide a good description of the
$\tau^-\to\pi^-\pi^0\nu_\tau$ spectral function for a squared $\pi\pi$
invariant mass up to about $s_{\rm max} \simeq 1.5$~GeV$^2$. Beyond these
energies, we propose a complementary expression for the form factor that
includes the effects of the excited states $\rho'$ and $\rho''$, matching
smoothly the previous one at $s \sim s_{\rm max}$.
Our result for the full form factor can be useful to improve the new
version of TAUOLA \cite{Shekhovtsova:2012ra}, which presently includes the
expressions obtained within a chiral Lagrangian framework with resonances
\cite{Jamin:2008qg, Arganda:2008, Dumm:2009kj, Dumm:2009va, Dumm:2012vb}.
This is important \cite{Actis:2010gg} not only for the proper simulation
of backgrounds and subsequent signal extraction at the more frequent tau
decay modes but also for the analysis of rare processes and the
searches of new physics \cite{Guo:2010ny}.

The article is organized as follows: in Section \ref{VFF low E} we obtain a
dispersive representation of $F_V^\pi(s)$ and discuss the inclusion of
isospin breaking corrections.
The model parameters are fitted to experimental data up to $s \simeq 1.5$
GeV$^2$, and our input for $\delta_1^1(s)$ in the elastic region is
confronted with present experimental values.
Then, in Section \ref{VFF Int E} we extend our parametrization of
$F_V^\pi(s)$ to higher energies, including the effective contribution of
excited resonances. The agreement with experimental data is shown.
In Section \ref{Low-E obs} we present the results for the low-energy
observables related to our subtraction constants.
Finally, in Section \ref{Concl} we state the conclusions of our analysis.

\section{Low energy description of $F_V^\pi(s)$}\label{VFF low E}

As stated, at very low energies the pion vector form factor is well
described by $\chi$PT. Let us first consider the limit of exact isospin
symmetry. At $\mathcal{O}(p^4)$, one has~\cite{Gasser:1984ux}
\begin{equation}\label{VFF ChPT}
 F_V^\pi(s)^{\chi\rm PT}\ =\ 1\, + \, \frac{2L_9^r(\mu)}{F_\pi^2}\, s\, -
 \,\frac{s}{96\pi^2F_\pi^2}
 \left[A_\pi(s,\mu^2)+\frac{1}{2}A_K(s,\mu^2)\right]\ ,
\end{equation}
where $L_9^r(\mu)$ is one of the renormalized low-energy coupling constants
in the chiral Lagrangian. The functions $A_P(s,\mu^2)$ are given by
\begin{equation} \label{A_P}
 A_P(s,\mu^2)\ =\ \log\frac{m_P^2}{\mu^2}\,+\,8\frac{m_P^2}{s}
 -\frac{5}{3}+\sigma_P^3(s)\, \log\left(
 \frac{\sigma_P(s)+1}{\sigma_P(s)-1}\right)\ ,
\end{equation}
where the phase space function $\sigma_P(s)$ reads
\begin{equation} \label{sigma_P}
 \sigma_P(s)=\sqrt{1-4\frac{m_P^2}{s}}\,.
\end{equation}
On the other hand, the computation of $F_V^\pi(s)$ from a chiral Lagrangian
that includes the lowest-lying vector meson multiplet as active resonance
fields yields~\cite{Ecker:1988te}
\begin{equation}
 F_V^\pi(s)\, = \, 1\, + \,\frac{F_V\, G_V}{F_\pi^2}\,\frac{s}{M_V^2-s}\ ,
\end{equation}
where $M_V=M_\rho$, and $F_V$ and $G_V$ measure the strength of the $\rho
V_\mu$ and $\rho\pi\pi$ couplings, respectively, $V_\mu$ being the quark
vector current. This tree level result corresponds to the leading term in
powers of $1/N_C$, and it is $\mathcal{O}(p^4)$ in the chiral expansion. If
the form factor is required to vanish in the limit of large $s$, then one
gets the relation $F_V G_V = F_\pi^2$, which yields
\begin{equation}\label{VMD}
 F_V^\pi(s)=\frac{M_\rho^2}{M_\rho^2-s}\ .
\end{equation}
Comparing with Eq.~(\ref{VFF ChPT}), the low energy
$\chi$PT coupling $L_9$ is predicted to be
\begin{equation}\label{L_9 large N_C}
 L_9=\frac{F_VG_V}{2M_\rho^2}=\frac{F_\pi^2}{2M_\rho^2}\simeq
7.2\cdot10^{-3}\ ,
\end{equation}
in very good agreement with the value obtained from phenomenology. This
shows explicitly that the $\rho(770)$ contribution is the dominant
physical effect in the vector form factor of the pion. Now, as stated in
Ref.~\cite{Guerrero:1997ku}, one can do better and match Eq.~(\ref{VMD})
to the $\mathcal{O}(p^4)$ $\chi$PT result in Eq.~(\ref{VFF ChPT}),
including the final state interactions encoded in the chiral loop
functions $A_P(s,\mu^2)$:
\begin{equation}
 F_V^\pi(s)=\frac{M_\rho^2}{M_\rho^2-s}-
 \frac{s}{96\pi^2F^2}\left[A_\pi(s)+\frac{1}{2}A_K(s)\right]\ .
 \label{fvresum}
\end{equation}
We omit from now on the explicit dependence on the $\mu$ scale, taking $\mu
= M_\rho$. The results do not depend significantly on changes in this scale.
From Eq.~(\ref{fvresum}), unitarity and analyticity constraints lead to the
Omn\`es exponentiation of the full loop function~\cite{Guerrero:1997ku},
\begin{equation}
F_V^\pi(s)\, = \,
\frac{M_\rho^2}{M_\rho^2-s}\,\exp\left\lbrace -\,\frac{s}{96\pi^2F^2}
\left[A_\pi(s)+\frac{1}{2}A_K(s)\right]\right\rbrace\ .
\end{equation}
In order to account for the resonance width, here one should not simply
replace $M^2-s$ by $M^2-s-iM\Gamma(s)$ in the effective propagator, since
this would double count $\Im m[A_P(s)]$ and analyticity would be violated at
$\mathcal{O}(p^6)$ in the chiral expansion. One could avoid the double
counting by shifting the imaginary part of the loop functions from the
exponential to the propagator~\cite{Guerrero:1997ku}, but still analyticity
would be lost. We follow instead a procedure similar to that proposed in
Ref.~\cite{Boito:2008fq} for the $K\pi$ form factor, in which unitarity and
analyticity are preserved. As a starting point we consider a form factor in
which the loop functions are resummed into the denominator,
\begin{eqnarray} \label{SU2formula}
\hspace{-.5cm} F_V^{\pi\,(0)}(s) & = & \frac{M_\rho^2}{M_\rho^2
\left[1+\frac{s}{96\pi^2 F_\pi^2}\left(A_\pi(s)+
\frac12 A_K(s)\right)\right]-s}\nonumber \\
& = & \frac{M_\rho^2}{M_\rho^2 \left[1+\frac{s}{96\pi^2 F_\pi^2}\Re e
\left(A_\pi(s) + \frac12 A_K(s)\right)\right]-s-i M_\rho
\Gamma_\rho(s)}\ ,
\end{eqnarray}
where we have defined the imaginary part of the denominator as $- M_\rho
\Gamma_\rho(s)$. The energy dependent width is thus given by
\begin{equation}
\Gamma_\rho(s)\, = \, -\frac{M_\rho\, s}{96\, \pi^2\, F_\pi^2}\Im m\left[A_\pi(s)
+\frac12 A_K(s)\right]\ ,
\end{equation}
and from Eq.~(\ref{A_P}) one has
\begin{equation}
 \Gamma_\rho(s) =  \frac{s\,M_\rho}{96\,\pi\,F_\pi^2}
 \left[\theta\left(s-4m_\pi^2\right)\,
 \sigma_\pi^3(s)+\frac{1}{2}\,\theta\left(s-4m_K^2\right)\,\sigma_K^3(s)\right]\
 ,
\label{Gammarho}
\end{equation}
which is in agreement with the result obtained from a chiral theory with
resonances~\cite{GomezDumm:2000fz} if one assumes the relation
$G_V=F_V/2$.

The form factor in Eq.~(\ref{SU2formula})
has the correct low-energy behaviour at ${\cal O}(p^4)$
\cite{Guerrero:1997ku} and leading ${\cal O}(p^6)$ contributions in $\chi$PT
\cite{Guerrero:1998hd}, and vanishes at short distances as expected from the
asymptotic behaviour ruled by QCD. As stated, the loop functions $A_P(s)$
contain the logarithmic corrections induced by final state interactions. Now
we take into account the fact that the two pion vector form factor is an
analytic function in the complex plane, except for a cut along the positive
real axis starting at the threshold for two pion production, $s_{\rm
thr}=4m_\pi^2$, where its imaginary part develops a discontinuity. From
unitarity it can be shown~\cite{Guerrero:1997ku,Pich:2001pj} that the form
factor satisfies an $n$-subtracted dispersion relation that involves the
scattering phase in the elastic region, for which experimental data are
available. In the case of $n$ subtractions at $s=0$, the dispersion relation
admits the well-known Omn\`es solution 
\begin{equation}
\label{omel}
F_V^\pi(s) \,=\, P_n(s) \exp \Biggl\{ \frac{s^n}{\pi}\!
\int\limits^\infty_{s_{\rm thr}}\!\!ds'\, \frac{\delta_1^1(s')}
{(s')^n(s'-s-i\epsilon)}\Biggr\} \ ,
\end{equation}
where
\begin{equation}
\label{Pns}
\log P_n(s) \,=\, \sum\limits_{k=0}^{n-1} \,\alpha_k \,
\frac{s^k}{k!}
\end{equation}
is the corresponding subtraction polynomial, and $\delta_1^1(s)$ is the
$I=1$, $J=1$ two-pseudoscalar scattering phase shift. The subtraction
constants $\alpha_k$ are given by
\begin{equation}
\alpha_k \, =\, \frac{d^k}{ds^k}\log F_V^\pi(s)\biggr|_{s=0} \ .
\end{equation}
At least one subtraction is required in Eq.~(\ref{omel}) to achieve
convergence. Here the first subtraction constant has been fixed from the
normalization $F_V^\pi(0) = 1$, which holds with good approximation in
view of the conservation of the vector current in the isospin symmetry
limit. On the other hand, in order to determine the scattering phase shift
$\delta_1^1(s)$ [to be used as input in Eq.~(\ref{omel})] we follow the
approach in Ref.~\cite{Boito:2008fq}, taking
\begin{equation}
\label{delta}
 \tan \delta_1^1(s) = \frac{\Im m F_V^{\pi(0)}(s)}{\Re e
F_V^{\pi(0)}(s)} \ ,
\end{equation}
where $F_V^{\pi(0)}(s)$ is given by Eq.~(\ref{SU2formula}). In this way, the
form factor in Eq.~(\ref{SU2formula}) trivially satisfies the Omn\'es
relation (\ref{omel}) for $n=1$ and $F_V^\pi(0)=1$. This form factor should
be adequate to reproduce the experimental observations in the low energy
limit, since by construction it matches $\chi$PT results. However, beyond
this limit one would not expect a sufficiently accurate description of the
data. Fortunately, the analyticity properties of $F_V^\pi(s)$ allow to
increase accuracy by considering more subtractions in Eq.~(\ref{omel}): for
higher $n$, the weight of the dispersive integral at large energies gets
reduced, and the corresponding information is translated to the subtraction
constants~\cite{Pich:2001pj}, which can be taken as unknown parameters. In
addition, some approach has to be used to deal with the phase shift beyond
the inelastic two-kaon threshold, where the contribution of the dispersive
integral is in general still relevant and Eq.~(\ref{omel}) is no longer
valid (in fact, this happens already at the four-pion threshold, but higher
multiplicity intermediate states are expected to be phase space suppressed).
The goal is to obtain a form factor that leads to a satisfactory description
of the available data, considering just a few subtractions and a
phenomenologically adequate elastic phase shift.

On the basis of the previous discussion, we have carried out fits of
$F_V^\pi(s)$ to Belle data from $\tau$ decays. We find that a good
description of the data can be obtained with $n=3$ subtractions, i.e.~taking
\begin{equation}\label{FV_3_subtractions}
 F_V^\pi(s) \,=\,\exp \Biggl[ \alpha_1\, s\,+\,\frac{\alpha_2}{2}\,
s^2\,+\,\frac{s^3}{\pi}\! \int^\infty_{s_{\rm thr}}\!\!ds'\,
\frac{\delta_1^1(s')} {(s')^3(s'-s-i\epsilon)}\Biggr] \ .
\end{equation}
In this form factor we have four parameters, namely the subtraction
constants $\alpha_{1,2}$, and the parameters $M_\rho$ and $F_\pi$ that
determine the phase shift $\delta_1^1$ according to
Eqs.~(\ref{SU2formula}) and (\ref{delta}). In addition, in order to deal
with the phase shift in the large energy region, we have distinguished two
intermediate (squared) energies $s_1$ and $s_2$. The former is defined as
the limit up to which we consider Eq.~(\ref{delta}) to be a reliable
description of the phase shift. As stated, we expect this value to be of
the order of the inelastic two-kaon threshold, $s_1\simeq 4M_K^2$, or
alternatively we can consider the limit $s_1\simeq
(M_\rho+\Gamma_\rho)^2$, where the effect of the $\rho$ resonance should
dominate. In any case one expects $s_1$ to be about 1 GeV$^2$. As we show
below, this will be supported {\em a posteriori} by the good agreement
between our predictions and the experimental data for $\delta_1^1$ quoted
some time ago in
Refs.~\cite{Hyams:1973zf,Estabrooks:1974vu,Protopopescu:1973sh}. Beyond
$s_1$, the dispersive integral in Eq.~(\ref{FV_3_subtractions}) should be
affected not only by inelastic contributions but also by the presence of
excited resonance states. The other point, $s_2$, indicates the energy at
which we assume that the phase shift saturates its asymptotic value
$\delta_1^1(s\to\infty)=\pi$, corresponding to the existence of a single
narrow resonance~\cite{MartinSpearman, Leutwyler:2002hm,
RuizArriola:2008sq}\footnote{The asymptotic limit of the phase shift
obtained from Eq.~(\ref{delta}) slightly deviates from this value, owing
to the linear growth of $\Gamma_\rho$ with $s$.}. Here we take $s_2\simeq
M_\tau^2$. In the intermediate region, $s_1\leq s \leq s_2$, we assume for
simplicity a linear behaviour of $\delta_1^1$ with $s$. In order to take
into account the uncertainties arising from these assumptions, when
performing our fits we have considered possible variations of the values
of $s_1$ and $s_2$, and of the upper integration limit $s_\infty$, which
is usually taken to be in the range $[2.25,\infty]$
GeV$^2$~\cite{Pich:2001pj, Boito:2008fq, Boito:2010me}. The corresponding
effects on our results have been taken as part of the systematic error of
our theoretical approach.

Another aspect to be taken into account is the effect of isospin violating
corrections to the pion vector form factor. In general one has to
distinguish between the neutral and charged pion vector form factors, the
latter being defined by Eq.~(\ref{FF definition}). The corrections can be
expanded in powers of the quark mass difference and the electromagnetic
coupling, in addition to the chiral counting. At the leading order, the
spectral function for the decay $\tau^-\to\pi^-\pi^0\nu_\tau$ can be written
as~\cite{Cirigliano:2001er}
\begin{eqnarray}
\frac{d \Gamma(\tau^-\to\pi^-\pi^0\nu_\tau)}{ds} & = &
\frac{G_F^2\, m_\tau^3}{384\,\pi^3}\, S_{EW}\, |V_{ud}|^2 \,
\left(1-\frac{s}{M_\tau^2}\right)^2\left(1+\frac{2s}{M_\tau^2}\right)
\nonumber \\
& & \lambda^{3/2}
\left(1,\frac{m_{\pi^0}^2}{s},\frac{m_{\pi^+}^2}{s}\right)
\,|f_+(s)|^2\,G_{EM}(s) \ ,
\label{spectral}
\end{eqnarray}
where $S_{EW} = 1+(\alpha/\pi) \log(M_Z^2/M_\tau^2)$ includes the dominant
short-distance electroweak corrections, and the factor $G_{EM}(s)$ arises
from the contribution of electromagnetic loops. In the isospin limit one has
$S_{EW} = G_{EM} = 1$, $\lambda^{1/2}(1,m_{\pi}^2/s,m_{\pi}^2/s) =
\sigma_\pi(s)$, and the form factor in Eq.~(\ref{spectral}) reduces to the
pion vector form factor, $f_+(s) = F_V^\pi(s)$. The dominant isospin
breaking effect in $F_V^\pi(s)$ is that arising from phase space, i.e., from
considering different masses for the charged and neutral pions and kaons in
the loop functions. Thus one has to replace the functions $A_\pi(s)$ and
$A_K(s)$ by $A_{\pi^-\pi^0}(s)$ and $A_{K^-K^0}(s)$, respectively. Explicit
expressions for these functions are given in Appendix A. In this way,
following the same steps that lead to Eq.~(\ref{FV_3_subtractions}) we can
obtain a dispersion relation for the charged pion vector form factor,
$F_V^{\pi+}(s)$.
In addition, in $f_+(s)$ one has to take into account a local
electromagnetic correction $f^{\mathrm{elm}}_{\mathrm{local}}$, which
contributes as an additional term in the decay
amplitude~\cite{Cirigliano:2001er}. One has then
\begin{equation}
f_+(s) \, = \, F_V^{\pi+}(s) \, + \, f^{\mathrm{elm}}_{\mathrm{local}} \ .
\end{equation}
This local electromagnetic correction is given by~\cite{Cirigliano:2001er}
\begin{equation}
 f^{\mathrm{elm}}_{\mathrm{local}} =
\frac{\alpha}{4\pi}\left( - \frac{3}{2} - \frac{1}{2}
\log\frac{M_\tau^2}{\mu^2} - \log\frac{m_\pi^2}{\mu^2} +
2\log\frac{M_\tau^2}{M_\rho^2} - X(\mu)\right)\ ,
\label{felm}
\end{equation}
where the scale dependence in the last term cancels that in the logarithms.
At the scale $\mu = M_\rho$, $X(\mu)$ is estimated to be between $-2.5$ and
$4.5$~\cite{Cirigliano:2001er}. Finally, the loop correction $G_{EM}(s)$ has
been computed (including resonance contributions) in
Refs.~\cite{Cirigliano:2002pv} and \cite{FloresBaez:2006gf}. We notice that
this correction has not been taken into account in the extraction of the
form factor carried out by the Belle Collaboration in Refs.~\cite{Belle,
Hiyashii} (it has been included for the analysis of the muon anomalous
magnetic moment, where isospin breaking effects represent a central subject
of interest).

In order to incorporate the effect of isospin corrections and evaluate its
significance, we have carried out our fits for the tau decay data
considering three different situations:
\begin{itemize}
 \item I) The limit of exact isospin symmetry, in which $f_+(s) =
F_V^\pi(s)$, where the form factor is given by
Eq.~(\ref{FV_3_subtractions}).
 \item II) The inclusion of isospin breaking corrections at the
level of kinematics, i.e.~considering different masses for the charged and
neutral particles in the loop functions and the kinematical factors in
Eq.~(\ref{spectral}) (this would correspond to Belle's analysis~\cite{Belle}
of the $\tau^-\to\pi^-\pi^0\nu_\tau$ decay width).
 \item III) The inclusion of all lowest order isospin breaking corrections, as
in Eqs.~(\ref{spectral}-\ref{felm}). For the factor $G_{EM}(s)$ we
have considered here the analysis in Ref.~\cite{FloresBaez:2006gf}.
\end{itemize}

As a general result, it is found that we obtain a good fit to Belle
data~\cite{Belle} for $s\lesssim 1.5$ GeV$^2$. Our fits have been carried
out with the MINUIT package taking the first 30 points ($s_{\rm max} =
1.525$ GeV$^2$, with a bin width of $0.05$ GeV$^2$). The results are shown
in Table \ref{Tab:1}.
First of all, it is worth to notice that ---even with just a few input
parameters--- the theoretical curve is able to fit the very precise set of
experimental data with a $\chi^2/dof$ value close to unity. Hence, it is
seen that within this energy range the data can be described without the
inclusion of higher resonant states in the theoretical scheme. On the
other hand, it is found that the effect of isospin breaking corrections on
the fitted parameters is below the 2\% level.

\begin{table*}[h!]
 \begin{center}
\begin{tabular}{|c||c|c|c|}
\hline
 & Fit value (I) & Fit value (II) & Fit value (III)\\
\hline
$M_\rho\,[{\rm GeV}]$& $0.8430(5)(17)$ & $0.8427(5)(14)$ & $0.8426(5)(20)$\\
$F_\pi\,[{\rm GeV}]$ & $0.0901(2)(5)$& $0.0902(2)(4)$& $0.0906(2)(4)$\\
$\alpha_1\,[{\rm GeV}^{-2}]$& $1.87(1)(3)$ & $1.87(1)(3)$ & $1.81(1)(2)$\\
$\alpha_2\,[{\rm GeV}^{-4}]$& $4.29(1)(7)$ & $4.31(1)(7)$ & $4.40(1)(6)$\\
\hline
$\chi^2/dof$ & $1.37$ & $1.37$ & $1.55$\\
\hline
$\Gamma_{\rho}(M_\rho^2)\,[{\rm GeV}]$ & $0.206(1)(3)$ & $0.206(1)(3)$ & $0.204(1)(3)$\\
\hline
\end{tabular}
\caption{\small{Results of our fits. The first and second numbers in
brackets correspond to the statistic and theoretical systematic errors,
respectively. $\Gamma_{\rho}(M_\rho^2$) is obtained using the fitted
values of $M_\rho$ and $F_\pi$ and is given only for reference.}} \label{Tab:1}
\end{center}
\end{table*}

Regarding the errors in the fitted parameters, we have quoted separately
those arising from the fit and the systematic errors coming from the
theoretical approach. The latter are basically due to the uncertainties in
the energy range to be fitted, the number of subtractions considered, and
the values of $s_1$, $s_2$ and $s_\infty$ in the dispersive integral. In
order to have an estimation of the effect of these uncertainties we have
considered the fits for $s_{\rm max}$ in the range $[1.325,1.525]$
GeV$^2$, 2 to 4 subtractions, and $s_1$, $s_2$, $s_\infty$ in the ranges
$[0.95,1.1]$\footnote{Best fits are obtained in all cases for $s_1\simeq
0.98$ GeV$^2$, in agreement with theoretical expectations.},
$[M_\tau^2,\infty]$ and $[2.25,\infty]$ GeV$^2$, respectively. The
corresponding results have been quoted in the second brackets in Table
\ref{Tab:1}, while the numbers in the first brackets stand for the
statistical errors arising from the fit.
It is found that $M_\rho$ and $F_\pi$ appear to be anticorrelated, and the
same happens with the parameters $\alpha_1$ and $\alpha_2$.

{}From the table it is seen that the central values of $F_\pi$ obtained
from the fit are about two percent below the value of 92.2 MeV quoted by
the PDG~\cite{Beringer:1900zz}. The difference can be attributed to
further theoretical uncertainties, mainly arising from the effect of
higher order terms in the large $N_C$ expansion. This includes corrections
to the relations $G_V = F_V/2$ and $G_V F_V = F_\pi^2$, which
have been used for the matching between the form factors obtained within
the low energy $\chi$PT theory and the chiral theory with resonances. In
fact, as already pointed out in Refs.~\cite{Pich:2001pj, Boito:2008fq},
the energy-dependent width given by the imaginary part of the loop
function with $F_\pi = 92.2$ MeV is not adequately normalized so as to
reproduce both the experimental data on $\pi\pi$ and $K\pi$ tau decay
channels. Concerning the properties of the $\rho(770)$ resonance, in order
to obtain the corresponding physical mass and width one should compute the
position of the pole of the pion vector form factor in the complex $s$
plane, say $s_\mathrm{pole}$. One has
\begin{equation} \label{spole}
\sqrt{s_\mathrm{pole}}\, =
\,M_\rho^\mathrm{pole}-\frac{i}{2}\Gamma_\rho^\mathrm{pole}\ .
\end{equation}
Unfortunately, $s_\mathrm{pole}$ cannot be obtained directly from the
expression for the pion vector form factor in Eq.~(\ref{FV_3_subtractions}),
since in general the complex variable $s$ in the dispersion relation is not
in the same Riemann sheet in which the pole is located. In order to deal
with this difficulty, one possible procedure is to make use of one-pole
Pad\'e approximants $P_1^N(s;s_0)$, defined by
\begin{equation}\label{P_1^N}
P_1^N(s;s_0)=\sum_{k=0}^{N-1}a_K(s-s_0)^k +
\frac{a_N(s-s_0)^N}{1-\frac{a_{N+1}}{a_N}(s-s_0)}\ .
\end{equation}
In general, if one assumes that a complex function $F(s)$ is analytical in a
disk around some point $s_0$ except at a point $s_\mathrm{pole}$, where it
has a single pole, then the de Montessus de Ballore's theorem~\cite{MdeB}
states that the sequence of one-pole Pad\'e approximants $P_1^N(s;s_0)$
converges to $F(s)$ in any compact subset of the disk excluding the pole.
Hence, the Pad\'e pole $z_p = s_0 + a_N/a_{N+1}$ of $P_1^N(s;s_0)$ converges
to $s_\mathrm{pole}$ for $N\to\infty$. The application of this method for
the analysis of resonance poles has been previously considered in
Refs.~\cite{Masjuan:2010ck}, where details can be found. In our case we have
approximated the form factor $F_V^\pi(s)$ with a function of the type of
that in Eq.~(\ref{P_1^N}), taking $s_0 = (M_0 - i\Gamma_0/2)^2$, with $M_0 =
0.77$~GeV, $\Gamma_0 = 0.15$~GeV. The coefficients $a_K$, $K=1,\dots ,N+1$
($a_0 = 1$ owing to vector current conservation) have been determined from a
fit to a set of values of $|F_V^\pi (s)|$ and $\delta_1^1(s)$ obtained from
our dispersive representation, Eq.~(\ref{FV_3_subtractions}), between the
first and second production thresholds. From the results of this fit, taking
Pad\'e approximants with $N=5$ and $N=6$ coefficients, we find
\begin{eqnarray}
& & \hspace*{-1cm}
M_\rho^\mathrm{pole}\,=\,(759\pm 2)\,\mathrm{MeV}\ ,
 \quad \Gamma_{\rho}^\mathrm{pole}\,=(146\pm 6)\,\mathrm{MeV}\ \quad
 {\rm (Fit\ I)} \ ;\nonumber\\
& & \hspace*{-1cm} M_\rho^\mathrm{pole}\,=\,(760\pm 2)\,\mathrm{MeV}\ ,
 \quad \Gamma_{\rho}^\mathrm{pole}\,=(147\pm 6)\,\mathrm{MeV}\ \quad
 {\rm (Fit\ III)} \ .
 \label{Pade Pole values}
\end{eqnarray}
This turns out to be our best determination of $s_\mathrm{pole}$. One can
still increase $N$ and get a better fit of the data set, but given the
larger number of parameters, the errors of $a_N$ and $a_{N+1}$ become also
larger. As expected, the values in Eq.~(\ref{Pade Pole values}) are not
modified either if we take a different input for $s_0$ or if we increase the
number of values of $|F_V^\pi (s)|$ and $\delta_1^1(s)$ to be fitted.

For comparison, we have also analyzed the results for the pole mass and
width of the $\rho$ meson corresponding to the parametrization proposed time
ago by Gounaris and Sakurai (GS)~\cite{Gounaris:1968mw}, which has been used
in the fits carried out by the Belle
Collaboration~\cite{Achasov:2011ra}\footnote{Details of this parametrization
are given in the next section, see Eq.~(\ref{FV BWs}) and below.}. The
results obtained by Belle using the normalization $F_V^\pi(0)=1$ yield the
parameter values $M_\rho^{\rm (GS)}=(774.6\pm0.5)$~MeV, $\Gamma_\rho^{\rm
(GS)}=(148.1\pm1.7)$~MeV. Taking into account the prescriptions in
Refs.~\cite{Bhattacharya:1991gr, Bernicha:1994re, Escribano:2002iv} to deal
with the cuts in the complex functions entering the GS form factor [see
Eqs.~(\ref{gs1}, \ref{BWs}) below] these parameters correspond to
\begin{equation}\label{Pole values GS}
 M_\rho^\mathrm{pole}\,=\,(760.9\pm 0.6)\,\mathrm{MeV}\ ,
 \quad \Gamma_{\rho}^\mathrm{pole}\,=(142.2\pm 1.6)\,\mathrm{MeV}\ .
\end{equation}
The results in Eqs.~(\ref{Pade Pole values}) and (\ref{Pole values GS}) are
consistent with each other and somewhat different from the average values
for the $\rho$ mass and width quoted by the PDG, namely $M_\rho = 775.49\pm
0.34$~MeV and $\Gamma_\rho = 149.1\pm 0.8$~MeV~\cite{Beringer:1900zz}. In
fact, the PDG values correspond to the parameters appearing in
phenomenological amplitudes where the resonances are introduced through BW
functions (as e.g.\ $M_\rho^{\rm (GS)}$ and $\Gamma_\rho^{\rm (GS)}$), hence
they are strongly model dependent\footnote{This has been pointed out in
Refs.~\cite{Boito:2008fq, Boito:2010me} for the case of the $K^\star(892)$
resonance, analyzed in the context of $\tau^-\to(K\pi)^-\nu_\tau$ decays.}.
Alternatively, one can take the pole mass and width as the relevant
resonance properties. We consider the agreement between the results in
Eqs.~(\ref{Pade Pole values}) and (\ref{Pole values GS}) as a check of
consistency, in the sense that one expects the pole parameters to be
essentially model independent. For comparison, in Table~\ref{Tab:Pole
Values} we show other determinations of the pole mass and width quoted in
the literature. The results for $M_\rho^\mathrm{pole}$ and
$\Gamma_\rho^\mathrm{pole}$ obtained either from our dispersive approach or
from the simple GS parametrization are found to be in good agreement with
the average value of these determinations. It is seen that the errors in
Eqs.~(\ref{Pole values GS}) are smaller than those in our results, since the
former were obtained from a direct fit to experimental data, while in our
determination one has an additional uncertainty introduced by the Pad\'e
approximants. However, the GS parametrization represents just a simple
ad-hoc description of the underlying dynamics, thus it implicitly includes a
theoretical systematic error which is hardly estimable.

\begin{table*}[h!]
 \begin{center}
\begin{tabular}{|c|c|c|c|c|}
\hline
\small{Reference} & \small{$M_\rho^\mathrm{pole}$} & \small{$\Gamma_{\rho}^\mathrm{pole}$} & \small{Data} & \small{Analysis}\\
\hline
\small{Sanz-Cillero {\em et al.}} \cite{SanzCillero:2002bs}& \small{$764.1^{+4.8}_{-3.7}$} & \small{$148.2^{+2.5}_{-6.2}$}& \small{$\tau\,\&\, e^+e^-$} & \small{DSE}\\
\small{Ananthanarayan {\em et al.}} \cite{Ananthanarayan:2000ht}& \small{$762.5\pm2$} & \small{$142\pm7$}&\small{$\pi\pi\to\pi\pi$} & \small{RE}\\
\small{Feuillat {\em et al.}} \cite{Feuillat:2000ch}& \small{$758.3\pm5.4$}& \small{$145.1\pm6.3$}& \small{$\tau\,\&\, e^+e^-$} & \small{$S$MA}\\
\small{Pel\'aez} \cite{Pelaez:2004xp}& \small{$754\pm18$}& \small{$148\pm20$}& \small{$\pi\pi\to\pi\pi$} & \small{U$\chi$PT}\\
\small{Zhou {\em et al.}} \cite{Zhou:2004ms}& \small{$763.0\pm0.2$} & \small{$139.0\pm0.5$}& \small{$\pi\pi\to\pi\pi$} & \small{$\chi$U}\\
\small{Masjuan {\em et al.}} \cite{Masjuan:2010ck}& \small{$763.7\pm1.2$} & \small{$144\pm3$}& \small{$\tau$} &\small{RA}\\
\small{Results from our fit I} & \small{$759\pm 2$} & \small{$146\pm 6$}& \small{$\tau$} & \small{DR}\\
\small{Results from our fit III} & \small{$760\pm 2$} & \small{$147\pm 6$}& \small{$\tau$} & \small{DR}\\
\small{Results from GS model} & \small{$760.9\pm 0.6$} & \small{$142.2\pm 1.6$}& \small{$\tau$} & \small{GS}\\
\hline
\end{tabular}
\caption{\small{Comparison between different results for the pole mass and
width of the $\rho(770)$ meson (values are in MeV). Abbreviations for the
type of analysis carried out are DSE: Dyson-Schwinger equations; RE: Roy
equations; $S$MA: $S$ matrix approach; U$\chi$PT: Unitarized Chiral
Perturbation Theory; $\chi$U: Chiral unitarization; RA: Rational
approximants; DR: Dispersive representation; GS: Gounaris-Sakurai
parametrization.}}\label{Tab:Pole Values}
\end{center}
\end{table*}

\begin{figure}[!h]
\begin{center}
\vspace*{1.25cm}
\includegraphics[scale=0.45,angle=-90]{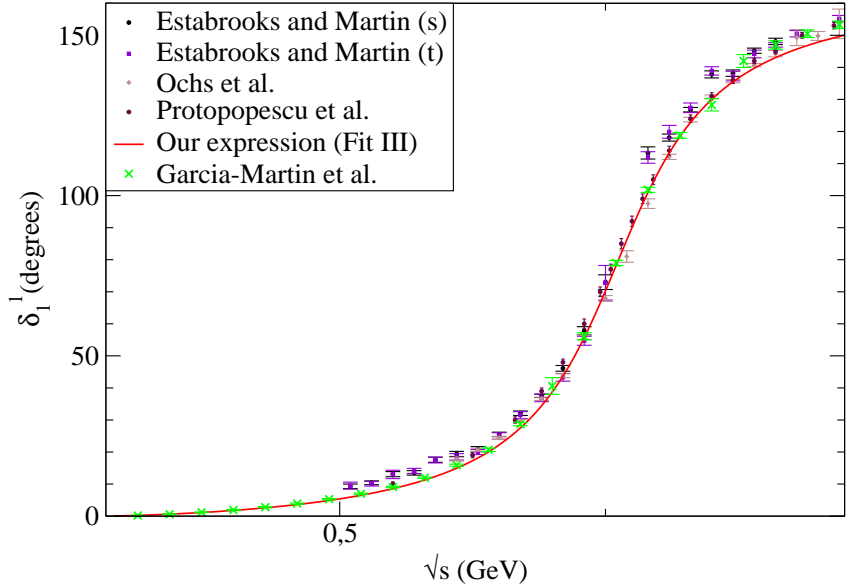}
\caption[]{\label{fig:1} \small{Two-pion phase shift $\delta_1^1$ as
function of the $\pi\pi$ invariant mass squared. Our theoretical
expression (red curve) is shown to be in good agreement with experimental
data (from Ochs \textit{et al}. \cite{Hyams:1973zf}, Estabrooks and Martin
\cite{Estabrooks:1974vu} in the $s$ and $t$ channels and Protopopescu
\textit{et al.} \cite{Protopopescu:1973sh}) up to the opening of the
two-kaon threshold, $s_1\simeq 1$~GeV$^2$. At very low energies, where no
data are available, our prediction agrees with the results of
Garc\'{\i}a-Mart\'{\i}n {\em et al.}~\cite{GarciaMartin:2011cn}.}}
\end{center}
\end{figure}

As a further check of consistency, we can compare the phase of the form
factor obtained from our fit with present experimental data on
$\delta_1^1(s)$ from Ochs \textit{et al.}~\cite{Hyams:1973zf}, Estabrooks
and Martin~\cite{Estabrooks:1974vu} and Protopopescu \textit{et
al.}~\cite{Protopopescu:1973sh}. The results are shown in Fig.~\ref{fig:1}.
We quote the data from threshold up to $s_1\simeq 1$~GeV$^2$, i.e.~the
region in which the phase of $F_V^{\pi\,(0)}(s)$ has been used as input for
the dispersive integral. In general it is seen that the agreement is very
good. It is remarkable, however, that our predictions are somewhat below the
data in the region of very low energies. On the other hand, as shown in the
figure, our results in that region are in very good agreement with those
recently obtained in Ref.~\cite{GarciaMartin:2011cn} using once-subtracted
Roy-like equations \cite{Roy:1971tc}. Within errors, we also find agreement
with the values obtained supplementing Roy equations with chiral symmetry
constraints \cite{Colangelo:2001df, Caprini:2011ky}. This discrepancy
between theory and experiment in the very low energy region would deserve
further tests, taking into account that the experimental information
corresponds to rather old measurements.

Another result that can be derived from our analysis is the value of the
point $s_{\pi/2}$, satisfying $\delta_1^1(s_{\pi/2})=\pi/2$. This point is
often used for the definition of the so-called peak mass or visible mass
(since it can be directly extracted from experimental data). According to
our fits, this value is rather stable with respect to changes in the above
mentioned systematic sources of theoretical errors. We find
\begin{equation}\label{peak mass}
 \sqrt{s_{\pi/2}}\,=\,(775.0\pm 0.2)\,{\rm MeV}\ .
\end{equation}
This is in agreement with the result of Ref.~\cite{Ananthanarayan:2000ht},
$\sqrt{s_{\pi/2}}\,=\,(774\pm3)$ MeV, within a Roy equation analysis of
the different partial waves in $\pi\pi$ scattering. The peak mass has to
coincide with the pole mass~\cite{Escribano:2002iv} when the corresponding
resonance is sufficiently narrow, it lies far from other resonances, and
there is a negligible background (non-resonant) contribution.

Finally, we point out that our results for the subtraction constants
$\alpha_1$ and $\alpha_2$ are compatible with previous determinations,
although the errors are smaller owing to the high quality of present Belle
data. We come back to this issue in Section \ref{Low-E obs}.

\section{Intermediate energy description of $F_V^\pi(s)$}\label{VFF Int E}

The approach presented in Sect.~\ref{VFF low E} has been used to obtain a
theoretical description of the $\tau\to\pi\pi\nu_\tau$ spectral function
for $\pi\pi$ invariant masses up to $s_{\rm max} \simeq 1.5$~GeV$^2$.
Above these energies this description is not adequate, in fact, the
experimental data are compatible with the presence of excited resonances.
In order to complement the dispersive representation for the pion vector
form factor proposed in the previous Section, we propose for the energy
region $s_{\rm max} \lesssim s \leq M_\tau^2$ an effective form factor that
includes two excited states, namely the $\rho'(1465)$ and $\rho''(1700)$
resonances. We stress that the dispersive representation is insensitive to
the modelling of this ``intermediate'' energy region. The dynamics related
with the excited resonance multiplets is poorly known, therefore there are
no grounds to go beyond a phenomenological parametrization that includes
several unknown parameters. We extend the form factor in
Eq.~(\ref{SU2formula}) as~\cite{Roig:2011iv}
\begin{eqnarray}\label{3rhosdisp}
 F_V^\pi (s) & = & \frac{M_\rho^2 \, + \, (\alpha^\prime
 e^{i\phi^\prime} + \alpha^{\prime\prime} e^{i\phi^{\prime\prime}})\, s}
 {M_\rho^2\left[1+\frac{s}{96\pi^2
 F_\pi^2}\left(A_\pi(s)+\frac12 A_K(s)\right)\right]-s}
 \nonumber \\
& & -\frac{\alpha' e^{i \phi^\prime}\, s}{M_{\rho^\prime}^2
\left[1+s\,C_{\rho^\prime} A_\pi(s)\right]-s}
-\frac{\alpha^{\prime\prime} e^{i \phi^{\prime\prime}}\, s}
{M_{\rho^{\prime\prime}}^2
\left[1+s\,C_{\rho^{\prime\prime}} A_\pi(s)\right]-s}\ ,
\end{eqnarray}
where the constants $C_{\rho^\prime}$ and $C_{\rho^{\prime\prime}}$ are given
by
\begin{equation}
C_R \, = \, \frac{\Gamma_R}{\pi\, M_R^3\,\sigma_\pi^3(M_R^2)}\ .
\end{equation}
The resonance masses $M_{\rho',\rho''}$ and on-shell widths
$\Gamma_{\rho',\rho''}$ are free parameters of this effective form factor.
By construction, the off-shell widths of the excited resonances behave in
a similar way as the $\rho$ width~\cite{Dumm:2009va}, considering the
two-pion states as the dominant absorptive parts of the corresponding
self-energies:
\begin{equation}\label{eq:rho1width}
\Gamma_R(s) =
\Gamma_R\,\frac{s}{M_R^2}\,
\frac{\sigma_\pi^3(s)}{\sigma_\pi^3(M_R^2)}\,\theta (s - 4m_\pi^2)\ .
\end{equation}
In addition, the form factor includes the coefficients $\alpha'$ and
$\alpha^{\prime\prime}$, which measure the relative weight between the
contributions of different resonances, and the phases $\phi^\prime$ and
$\phi^{\prime\prime}$, which account for the corresponding interference.

\begin{figure}[!t]
\begin{center}
\vspace*{0.2cm}
\includegraphics[scale=0.45,angle=-90]{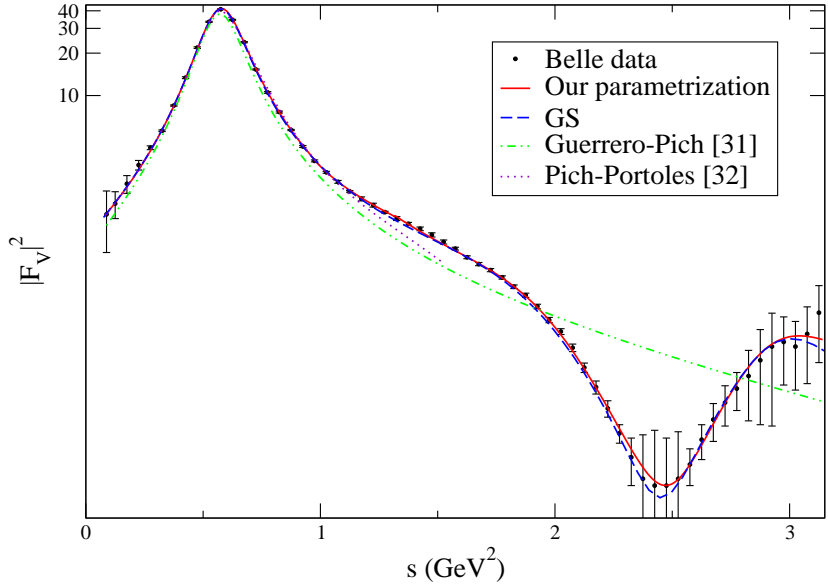}
\caption[]{\label{fig:2}
\small{Pion vector form factor $F_V^{\pi}(s)$ compared to Belle data
\cite{Belle} (black dots). Solid and dashed lines correspond to our
description and the GS parametrization, respectively. The dashed-dotted
curve stands for the result from Ref.~\cite{Guerrero:1997ku} (for
$M_\rho=775$ MeV), while the dotted line corresponds to the dispersive
representation in Ref.~\cite{Pich:2001pj} (for $\alpha_1=1.83$ GeV$^{-2}$,
$\alpha_2=4.32$ GeV$^{-4}$ and $M_\rho=774.2$ MeV).}}
\end{center}
\end{figure}

\begin{figure}[h!]
\centering
\subfigure{
\includegraphics[scale=0.24,angle=-90]{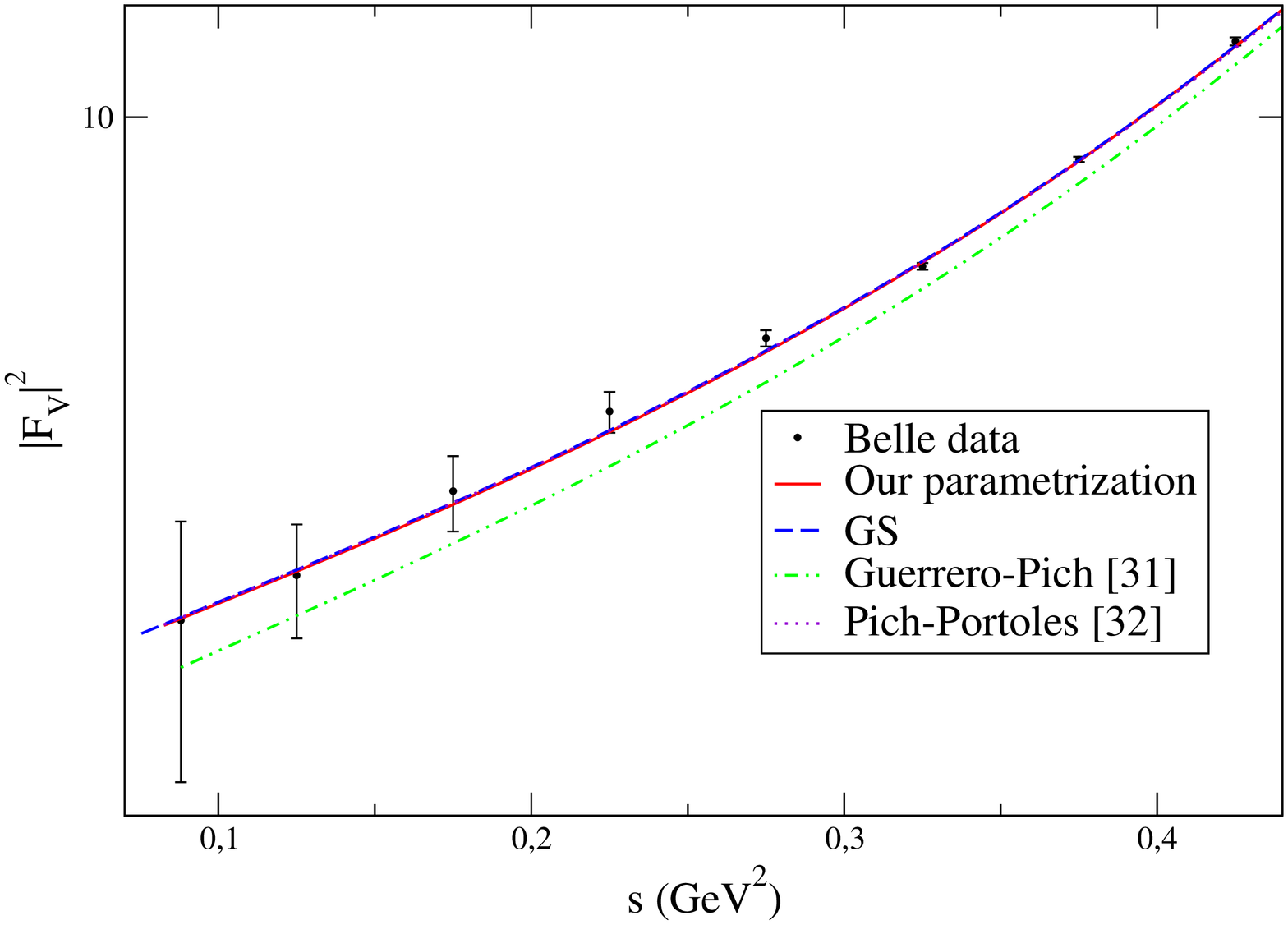}}
\hspace*{0.03cm}
\subfigure{
\includegraphics[scale=0.24,angle=-90]{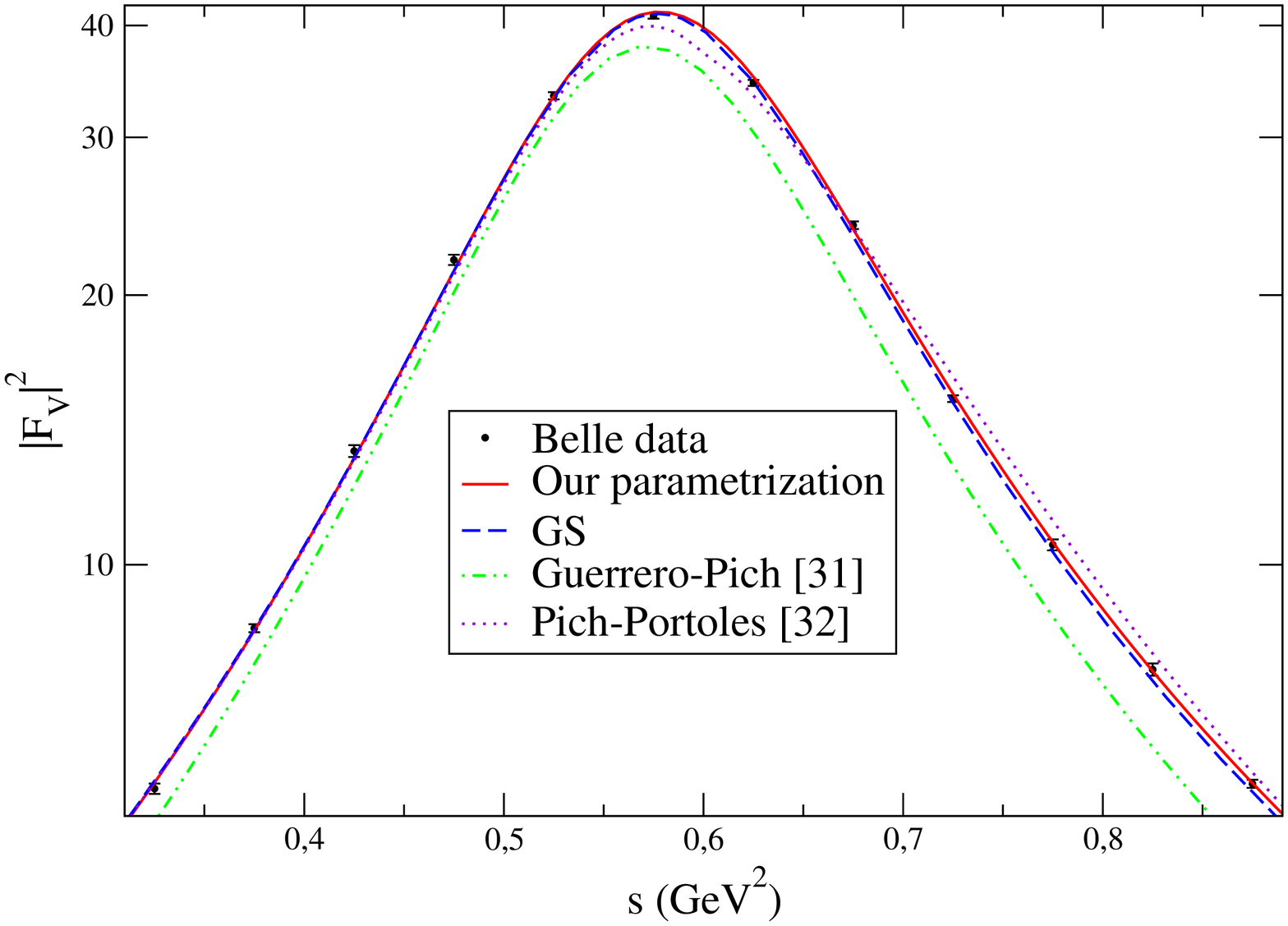}}
\caption{\small{Two close-ups of Fig.~\ref{fig:2} are displayed,
corresponding to the low-energy region (left panel) and the peak region
(right panel).} \label{fig:3}}
\end{figure}

Now the unknown parameters can be fitted to Belle data on the
$\tau\to\pi\pi\nu_\tau$ spectral function. The quality of the matching
between the phenomenological form factor in Eq.~(\ref{3rhosdisp}) and the
dispersive representation in Eq.~(\ref{SU2formula}) can serve as a test of
the consistency of our approach. The results of our fit for the resonance
parameters can be translated to the corresponding pole values, leading to
\begin{eqnarray}
& & M_{\rho'}^{\rm pole} = (1.44 \pm 0.08) \ {\rm GeV}
\ , \quad \Gamma_{\rho'}^{\rm pole} = (0.32 \pm 0.08) \ {\rm GeV}\ ,
\nonumber \\
& & M_{\rho''}^{\rm pole} = (1.72 \pm 0.09) \ {\rm GeV} \ ,
\quad \Gamma_{\rho''}^{\rm pole} =  (0.18 \pm 0.09) \ {\rm GeV} \ ,
\end{eqnarray}
in good agreement with the values quoted by the
PDG~\cite{Beringer:1900zz}. For the coefficients and phases we obtain
\begin{equation}
\begin{array}{ll}
\alpha^\prime = 0.08^{+0.03}_{-0.01} \ \quad
& \phi^\prime = 0.14^{+0.10}_{-0.08} \\
\rule{0cm}{.59cm}\alpha^{\prime\prime} = 0.03\pm 0.01 \ \quad
& \phi^{\prime\prime} = 3.14^{+0.50}_{-0.06} \ .
\end{array}
\end{equation}
Here, besides the statistical errors, we have included a systematic error
arising from the election of the initial value of the considered energy
range, say $s_0$. We have taken $s_0\in[1.3,1.55]$ GeV$^2$, and considered
fit results with $\chi^2/dof\leq 1$. Within this range we obtain a good
matching to the form factor in Eq.~(\ref{SU2formula}) at $s\simeq 1.35$
GeV$^2$. The fits are not significantly sensitive to the $\rho$ meson
parameters, which have been taken from the results in Table 1. Our final
curve for the pion vector form factor covering the full range of values from
threshold to $M_\tau^2$ is shown in Fig.~\ref{fig:2} (solid line). The
quality of the fits is reflected in the good agreement between our results
and the experimental data obtained by Belle, in particular in the low energy
region, where the latter are very precise. In addition, it can be seen that
the matching at $s\simeq 1.35$ GeV$^2$ is smooth, which supports the
consistency of the phenomenological description proposed for the
intermediate energy region. In order to appreciate the agreement with data
with more detail, two close-ups of Fig.~\ref{fig:2}, corresponding to the
low energy and the peak regions, are shown in Fig.~\ref{fig:3}.

It is worth to point out that the phenomenological form factor in
Eq.~(\ref{3rhosdisp}) is qualitatively similar to the GS
parametrization~\cite{Gounaris:1968mw} mentioned in the previous section.
Indeed, the GS form factor is built as a sum of Breit-Wigner-like
functions that keep a nontrivial real contribution in the corresponding
denominators:
\begin{equation}
\label{FV BWs}
F_V^{\pi {\rm (GS)}}(s) = \frac{1}{1 + \beta +\gamma }
      \left[BW_{\rho}^{\rm GS}(s) + \beta \, BW_{\rho^{\prime}}^{\rm GS}(s)
 +\gamma \, BW_{\rho^{\prime\prime}}^{\rm GS}(s)\right]\ ,
\label{gs1}
\end{equation}
where
\begin{equation}
\label{BWs}
BW_{R}^{\rm GS}(s) = \frac {M_{R}^2 \, ( 1 \, + \, d_R \,
\Gamma_{R}(s)/\sqrt{s}) }
       {(M_{R}^{2} - s) + f_R(s) - i M_R\, \Gamma_{R}(s)}\ ,
\end{equation}
and the coefficients $\beta$ and $\gamma$ are complex numbers. The
energy-dependent widths $\Gamma_R(s)$ are given, as in our approach, by
Eq.~(\ref{eq:rho1width}), while the expression for the (real) functions
$f_R(s)$ can be found in Ref.~\cite{Gounaris:1968mw}. The constants $d_R$
are chosen so that $BW_R^{\rm GS}(0) = 1$. As stated, this phenomenological
parametrization has been used in the fits carried out by the Belle
Collaboration~\cite{Achasov:2011ra}, allowing a quite successful description
of the data throughout the full spectrum. It is represented by the dashed
curve in Fig.~\ref{fig:2} (in the close-ups in Fig.~\ref{fig:3} our curve
and the GS curve overlap, and little differences can only be appreciated in
the peak region, where our curve shows a slightly better agreement with the
data). For comparison we also include in Figs.~\ref{fig:2} and \ref{fig:3}
the result obtained in Refs.~\cite{Guerrero:1997ku} and \cite{Pich:2001pj}.
The latter corresponds to a dispersive representation of the form factor in
the isospin limit, without the inclusion of excited resonant states (we have
refitted the parameters according to present Belle data).

\section{Low-energy observables}\label{Low-E obs}

On the basis of the theoretical approach presented in Sect.~\ref{VFF low
E} we can obtain the values of chiral low energy observables. If the
expansion of the pion vector form factor in powers of $s$ is parametrized
as
\begin{equation}
F_V^\pi(s)\, = \, 1 \, + \, \frac{1}{6}\,\left\langle
r^2\right\rangle^\pi_V \,s \, + \, c_V^\pi\, s^2 \, + \, d_V^\pi \, s^3\, + \, \dots \ ,
\end{equation}
{}from Eq.~(\ref{FV_3_subtractions}) one has
\begin{equation}
 \left\langle r^2\right\rangle^\pi_V=6\,\alpha_1\ ,\quad
 c_V^\pi=\frac{1}{2}\left(\alpha_2+\alpha_1^2\right)\ .
\end{equation}
Taking into account the results of our fit (case III, i.e.~including
isospin-breaking corrections), we obtain
\begin{equation}
 \left\langle r^2\right\rangle^\pi_V = 10.86 \pm 0.14\ {\rm GeV}^{-2}\
 ,\quad c_V^\pi=3.84 \pm 0.03\ {\rm GeV}^{-4} \ .
\end{equation}
These values are indeed in good agreement with almost all previous
determinations made by several authors within various chiral models, see
Refs.~\cite{Bijnens:1998fm,Bijnens:2002hp,
Pich:2001pj, Truong:1998yx,
de Troconiz:2004tr,Masjuan:2008fv,Guo:2008nc,Aoki:2009qn,
Ananthanarayan:2011xt,Colangelo:1996hs}. In order to go beyond the $s^2$
term in the expansion, one can make use of the general relation
\begin{equation}
 \alpha_k = \frac{k!}{\pi}\int_{4m_\pi^2}^{\infty}
 \mathrm{d}s^\prime\ \frac{\delta_1^1(s^\prime)}{{s^\prime}^{k+1}}\ ,
 \label{general}
\end{equation}
which allows to determine the subsequent subtraction constants in the
Omn\`es expression (\ref{omel}). In this way we obtain $\alpha_3 = 29.2
\pm 0.2$~GeV$^{-6}$, which leads to
\begin{equation}
 d_V^\pi\, = \,\frac{1}{6}\,(\alpha_3+3\alpha_1\alpha_2+\alpha_1^3)\, =
 \, 9.84\pm 0.05\ {\rm GeV}^{-6}\ .
\end{equation}
Previous evaluations for this observable have been carried out in
Refs.~\cite{Truong:1998yx} and \cite{Ananthanarayan:2011xt}, leading to
$9.70 \pm 0.40$~GeV$^{-6}$ and $10.18 \pm 0.27$~GeV$^{-6}$, respectively.
In order to check the consistency of our procedure we have also calculated
the constant $\alpha_2$ from the general relation (\ref{general}),
obtaining $\alpha_2=(3.7\pm0.2)$ GeV$^{-4}$, in reasonable agreement with
the result of the fit. Notice that, even if Eq.~(\ref{general}) is exact,
one can expect some deviation from the fitted value of $\alpha_2$ owing to
the ad-hoc treatment of the phase shift above the inelastic threshold in
our analysis of the form factor. In the case of $\alpha_1$ we cannot trust
the result from Eq.~(\ref{general}) since the slow convergence of the
integral provides a large weight to this high energy contribution.

\begin{table*}[h!]
 \begin{center}
\begin{tabular}{|c|c|c|}
\hline
Reference & $r^V_{V_1}(M_\rho)\times10^3$& $r^V_{V_2}(M_\rho)\times10^4$\\
\hline
VMD \cite{Bijnens:1998fm} & $-0.25$ & $2.6$\\
$\mathcal{O}(p^6)$ $\chi$PT \cite{Bijnens:1998fm} & $-0.68(26)$ & $1.50(44)$\\
Pich and Portol\'es~\cite{Pich:2001pj} & $-0.79(19)$ & $1.46(3)$\\
Our result & $-0.91(16)$ & $1.49(1)$\\
\hline
\end{tabular}
\caption{ \small{Counterterm combinations extracted from $\left\langle
r^2\right\rangle^\pi_V$ and $c_V^\pi$ in $\mathcal{O}(p^6)$
$\chi$PT.}} \label{Tab:3}
\end{center}
\end{table*}

In addition, the observables $\left\langle r^2\right\rangle^\pi_V$ and
$c_V^\pi$ can be related to two counterterm combinations in the
$\mathcal{O}(p^6)$ chiral Lagrangian, namely $r^r_{V_1}(M_\rho)$ and
$r^r_{V_2}(M_\rho)$~\cite{Bijnens:1998fm}, which are dominated by the
vector resonance contributions $r^V_{V_1}(M_\rho)$ and
$r^V_{V_2}(M_\rho)$. Our results for these quantities are quoted in
Table~\ref{Tab:3}, showing a good agreement with the values previously
obtained in Ref.~\cite{Pich:2001pj} and in the $\mathcal{O}(p^6)$ $\chi$PT
fit in Ref.~\cite{Bijnens:1998fm}. Within VMD these counterterms can be
determined by integrating out vector resonances in the framework of a
chiral effective theory~\cite{Ecker:1988te}. Considering just the
contribution of the $\rho$ meson resonance, within the Proca formalism one
gets~\cite{Bijnens:1998fm}
\begin{equation}\label{rV's Proca}
r^V_{V_1} = 2\sqrt{2}\frac{F_\pi^2}{M_V^2}f_\chi f_V \ ,
\quad r^V_{V_2} = \frac{F_\pi^2}{M_V^2}g_V f_V \ ,
\end{equation}
where $f_V$, $g_V$ and $f_\chi$ are effective couplings in the chiral
Lagrangian with resonances. Our results for $r^V_{V_1}$ and $r^V_{V_2}$
would lead then to the ratio $f_\chi/g_V = -2.1\pm 0.5$, far from the
phenomenological value $f_\chi/g_V\simeq -0.33$~\cite{Bijnens:1998fm}.
This indicates that the role of heavier resonances is crucial in order to
describe the $\mathcal{O}(p^6)$ vector driven contributions in $\chi$PT,
in agreement with Ref.~\cite{Pich:2001pj}.

\section{Conclusions}
\label{Concl} The high quality data on the pion vector form factor
obtained at flavour factories demands a correspondingly improved analysis
from the theoretical side. In order to describe these data keeping the
connection with the underlying strong interaction dynamics, one can take
profit of QCD symmetries to reproduce the data in the very low energy
domain, and make use of general properties of quantum field theory to
extend the analysis to higher energies. In this spirit, we have presented
a dispersive representation of the charged pion vector form factor that
fulfills the constraints imposed by analyticity and unitarity, and reduces
to the result obtained within $\chi$PT at low energies.

Our construction is based on the dispersion relation between the form
factor and the $\delta_1^1(s)$ phase shift of elastic $\pi\pi$ scattering.
The phase shift is obtained from the leading contribution arising in the
large-$N_C$ expansion including $\rho(770)$ exchange up the onset of
inelasticities, with the further assumption of a smooth growth up to the
asymptotic value. In this way we obtain a theoretical expression for the
form factor in terms of four parameters, namely $M_\rho$, $F_\pi$, and two
subtraction constants $\alpha_1$ and $\alpha_2$. The values of these
parameters have been determined by performing a fit to the very precise
Belle data on the $\tau^-\to\pi^-\pi^0\nu_\tau$ spectral function up to a
squared $\pi\pi$ invariant mass $s_{\rm max}\simeq 1.5$ GeV$^2$, leading
to the results quoted in Table~\ref{Tab:1}. It is seen that the effect of
isospin corrections on the parameters lies below the two percent level.
From these results we have determined the pole values of the $\rho$ mass
and width and the so-called visible or peak $\rho$ mass. We have also
obtained the values of low energy observables and compared the results
with those arising from chiral effective theories.

In addition, we have addressed the energy region $s\geq s_{\rm max}\simeq
1.5$ GeV$^2$, in which the inclusion of excited states is necessary to get
a proper description of $\tau^-\to\pi^-\pi^0\nu_\tau$ data. For this region we
have proposed a phenomenological expression for the form factor that takes
into account the presence of the resonances $\rho'$ and $\rho''$, assuming
that the effective propagators behave in a similar way as that of the
$\rho$ meson. This allows a good fit to the data, leading to values for
the $\rho'$ and $\rho''$ masses and widths similar to those quoted in
previous works. It is seen that the curves for the form factor obtained
for both energy regions match smoothly at $s \sim s_{\rm max}$.

As a conclusion, we have seen that the $\chi$PT results at low energies
supplemented with the leading contributions in the large-$N_C$ expansion
are able to provide the input to a dispersive representation of the pion
vector form factor which fulfills analyticity and unitarity. On this basis,
complemented with a phenomenological description in the high energy
region, we have shown that it is possible to reproduce the very precise
data on $\tau^-\to\pi^-\pi^0\nu_\tau$ decays throughout all the phase
space. This can be used as input to the new hadronic currents of the
TAUOLA Monte Carlo generator. On the other hand, our fits lead to the
parametrization of the charged pion vector form factor, thus the
comparison with a precise determination of the neutral form factor would
provide robust information on the $\pi\pi$ contribution to the muon
anomalous magnetic moment.

\section*{Acknowledgements}
We are grateful to J.J.~Sanz-Cillero and J.~Portol\'es for a critical
reading of our manuscript, and to M. Davier, R.~Escribano, G. L\'opez
Castro, B.~Moussallam and A. Pich for illuminating discussions.
P.R.~acknowledges J.~Bijnens and M.~Jamin for useful explanations. We also
thank H.\ Hayashii and D.\ Epifanov for their valuable information on the
analysis carried out by the Belle Collaboration, and G.\ Toledo for his
help on the evaluation of isospin-breaking corrections. This work has been
partially supported by the Spanish grants FPA2007-60323, FPA2011-25948 and
by the Spanish Consolider Ingenio 2010 Programme CPAN (CSD2007-00042). It
has also been founded in part by CONICET and ANPCyT (Argentina), under
grants PIP02495 and PICT-2011-0113, respectively.

\section*{Appendix}

The explicit form of the loop functions $A_{PQ}(s)$ can be obtained from
Ref.~\cite{Gasser:1984ux}. One has
\begin{equation} \label{A_PQ}
 A_{PQ}(s)\,=\,-\frac{192\,\pi^2\,\left[s\,M_{PQ}(s)-L_{PQ}(s)\right]}{s}\,,
\end{equation}
where $M_{PQ}(s)$ and $L_{PQ}(s)$ can be written in terms of new functions
$\Sigma_{PQ}$, $\Delta_{PQ}$, $k_{PQ}$, $\bar{J}_{PQ}$ and $\tilde{J}_{PQ}$
as
\begin{eqnarray}
M_{PQ}(s) & = & \frac{1}{12\,s}(s-2\,\Sigma_{PQ})\,\bar{J}_{PQ}(s)+
\frac{\Delta_{PQ}^2}{3\,s^2}\tilde{J}_{PQ}(s)-\frac{1}{6}k_{PQ}+
\frac{1}{288\,\pi^2} \nonumber \\
 L_{PQ}(s) & = & \frac{\Delta_{PQ}^2}{4\,s}\,\bar{J}_{PQ}(s)\ .
\end{eqnarray}
The new functions $\Sigma_{PQ}$ and $\Delta_{PQ}$ are defined by
$\Sigma_{PQ}=m_P^2+m_Q^2$, $\Delta_{PQ}=m_P^2-m_Q^2$, while $k_{PQ}$
includes the renormalization scale $\mu$:
\begin{equation}
k_{PQ} = \frac{F_\pi^2}{\Delta_{PQ}}(\mu_P-\mu_Q) \ ,
\end{equation}
where
\begin{equation}
\mu_P = \frac{m_P^2}{32\,\pi^2\,F_\pi^2} \log
\left(\frac{m_P^2}{\mu^2}\right) \
\end{equation}
(we have taken $\mu = M_\rho$, as in the isospin symmetric case). Finally,
the functions $\bar J_{PQ}$ and $\tilde J_{PQ}$ are given by
\begin{eqnarray}
\tilde{J}_{PQ}(s) & = & \bar{J}_{PQ}(s)-s \bar{J}'_{PQ}(0)\nonumber \\
\bar{J}_{PQ}(s) & = & \frac{1}{32\,\pi^2}\left[ 2 +
\left(\frac{\Delta_{PQ}}{s} - \frac{\Sigma_{PQ}}{\Delta_{PQ}}\right)\log
\left(\frac{m_Q^2}{m_P^2}\right) - \right. \nonumber \\
& & \frac{\nu}{s}\log
\left.\left(\frac{(s+\nu)^2-\Delta_{PQ}^2}{(s-\nu)^2-
\Delta_{PQ}^2}\right)\right] \ ,
\end{eqnarray}
where $\nu=\lambda^{1/2}(s,m_P^2,m_Q^2)$. We note finally that
\begin{equation}
 s\,\bar{J}'_{PQ}(0)=\frac{s}{32\pi^2}
 \left(
 \frac{\Sigma_{PQ}}{\Delta_{PQ}^2}
 + 2\frac{M_P^2 M_Q^2}{\Delta_{PQ}^3} \log
 \frac{M_Q^2}{M_P^2}\right) \ .
\end{equation}

\end{document}